\def\fm {\mathop{\hbox{fm}}}
\def\MeV {\mathop{\hbox{MeV}}}

\newcommand{\beq}{\begin{equation}}
\newcommand{\eeq}{\end{equation}}
\newcommand{\beqa}{\begin{eqnarray}}
\newcommand{\eeqa}{\end{eqnarray}}

\documentclass{PoS}

\title{Hadron electric polarizability -- finite volume corrections}

\ShortTitle{Hadron electric polarizability -- finite volume corrections}

\author{\speaker{Andrei Alexandru} \\
       George Washington University \\
       E-mail: \email{aalexan@gwu.edu}}

\author{Frank Lee\\
        George Washington University \\
        E-mail: \email{fxlee@gwu.edu}}

\abstract{We use the background field method to extract the polarizability for the neutral ``pion''.
In our previous study we found that the polarizability for this system is negative which is believed
to be a finite volume artifact. To address this issue, we carry out simulations for different lattice sizes
and we also look at the influence of the boundary conditions on these results. We find that for
pion masses lower than $400\MeV$ the polarizability remains negative even on larger lattices.
An infinite volume extrapolation is attempted, but the results are not conclusive due mainly to 
a lack of an analytical form for the finite volume corrections for this system.
}

\FullConference{The XXVIII International Symposium on Lattice Filed Theory\\
                June 14-19,2010\\
                Villasimius, Sardinia Italy}

\begin{document}

\section{Background and motivation}

In the presence of an electromagnetic field, hadron's charge distribution gets distorted and its energy
changes. Typical e-m fields induce small changes in the shape of hadrons and the energy shift is well 
parametrized by the leading order terms:
\beq
\Delta E = -\vec{p}\cdot\vec{E}-\vec{\mu}\cdot\vec{B}-\frac{1}{2}(\alpha E^2 + \beta B^2)+...
\label{eq:1}
\eeq
For hadrons, the electric polarizability vanishes, and the energy shift is proportional, at leading order,
with $E^{2}$. The constant of proportionality is related to the electric polarizability, $\alpha$, which is measured
in Compton scattering experiments. This quantity measures the strength of the electric dipole induced by
small electric fields.

On the lattice, the most common technique to compute electric (magnetic) polarizability is the background 
field method: a static electric (magnetic) field is introduced by minimally coupling it to the quarks' 
charges and the hadron mass shift induced by the field is measured. Using Eq.~\ref{eq:1} this energy 
shift is then used to compute the relevant polarizability. For electric polarizability, Eq.~\ref{eq:1} needs to
be modified to account for the fact that the lattice calculations are performed using Euclidean 
formulation~\cite{Alexandru:2008sj}. The background field method has been used to compute
the electric polarizabilities of neutral hadrons~\cite{Christensen:2004ca}, baryon magnetic
moments~\cite{Lee:2005ds} and polarizabilities~\cite{Lee:2005dq} and electric polarizabilities
for charged hadrons~\cite{Detmold:2009dx}. Most of these calculations were carried out using rather
heavy quarks. Since electric and magnetic polarizabilities are expected to change quite dramatically
as we approach the chiral limit, it is imperative to study the dependence on the quark mass of these
observables. Our first study towards this goal~\cite{Alexandru:2009id} showed that the dynamical 
and quench simulations produce rather similar result for pion masses as low as $400\MeV$, and that
the electric polarizability of the neutron starts raising significantly when $m_{\pi}< 400\MeV$. 

\begin{figure}[htbp]
\begin{center}
   \includegraphics[width=12cm]{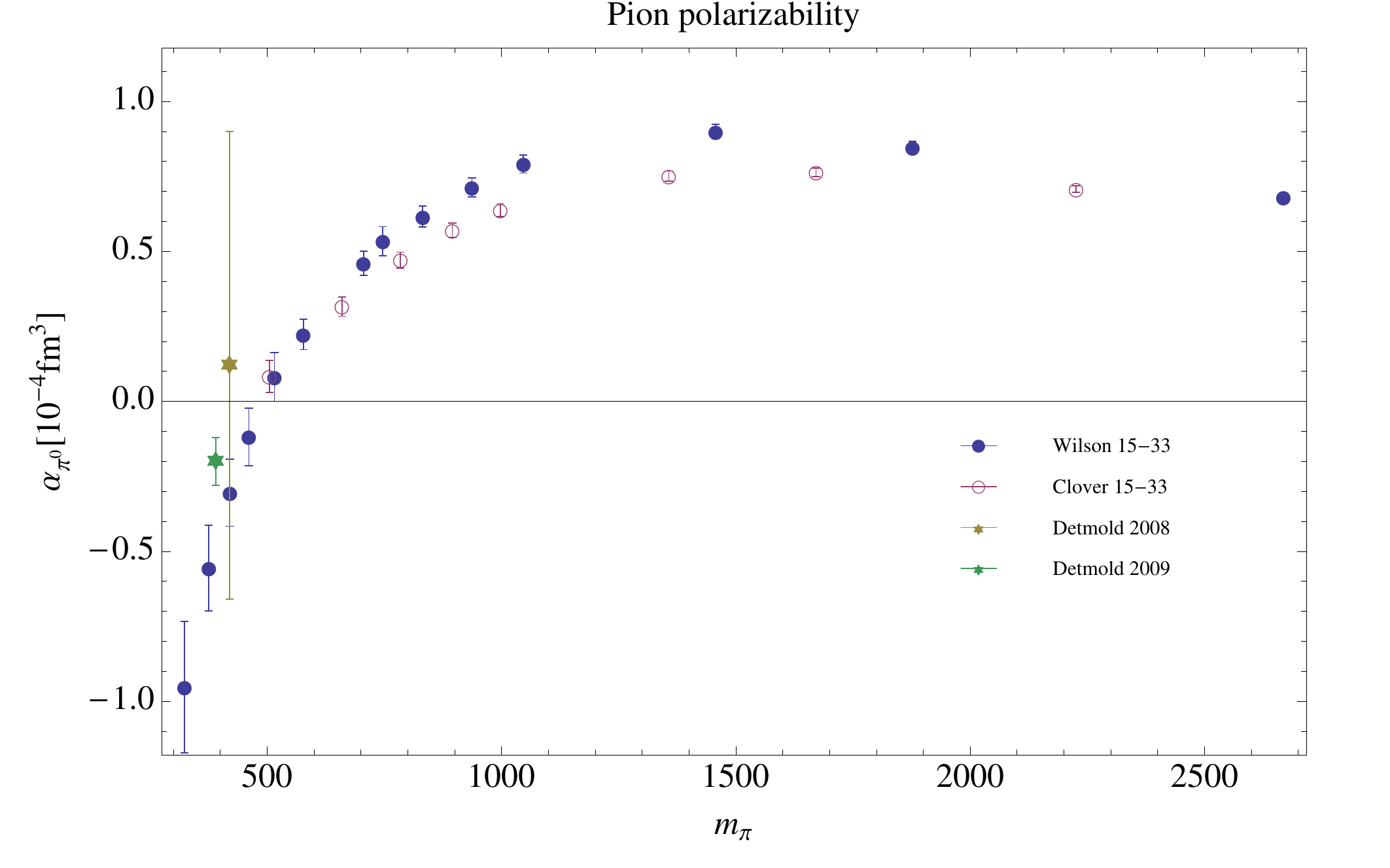} 
   \caption{Pion polarizability: the blue points are result calculated using Wilson fermions, 
   the empty point are clover results and the stars are the results 
   from~\cite{Detmold:2008xk,Detmold:2009dx}.
   \label{fig:1}}
\end{center}
\end{figure}

Another interesting result of our previous study is that the ``connected'' part of the neutral pion 
polarizability becomes more and more negative as the pion mass is lowered below $500\MeV$
(see Fig.~\ref{fig:1}). This has also been observed in calculations that employed dynamical 
configurations~\cite{Detmold:2009dx}. This is an interesting feature, not only because negative
electric polarizabilities are impossible to accommodate in a non-relativistic framework, but also
because the expectation from chiral effective field theory is that the ``connected'' polarizability
should be small and positive. The discrepancy is suspected to be due to finite volume corrections
which are expected to be large for these observables~\cite{Tiburzi:2008pa}.

To address this problem we decided to investigate the finite volume corrections to the polarizability.
We will focus here on the ``connected'' part of the polarizability for the neutral pion. We will first discuss
why the polarizability has rather large finite volume corrections and the advantages and disadvantages
of using different boundary conditions. We will then present the parameters used in our simulations and
our results.

\section{Background field method and boundary conditions}

The static electric field is introduced via minimal coupling:
\beq
D_\mu = \partial_\mu - i g G_\mu - i q A_\mu,
\eeq
where $G_{\mu}$ is the color field and $A_{\mu}$ the electromagnetic field.
On the lattice this amounts to a multiplicative factor for the links. This factor can be either real or
complex provided that the connection between the mass shift and polarizability is modified 
accordingly~\cite{Alexandru:2008sj}. In this study we introduce the electric field using a complex
phase, i.e. $U_\mu \rightarrow e^{-iqaA_\mu} U_\mu$. To introduce a constant electric field in the
$x$-direction we can choose $A_x = E t$ (the other components are set to zero) or $A_t=-E x$.
In this case the mass shift is $\Delta m = +\frac{1}{2} \alpha E^2$.

In order to create an uniform electric field, we have to carefully adjust the phase factor at lattice boundaries.
Both choices listed above will create a constant electric field at the interior points of the lattice. However,
the choice $A_{x} = E t$ will have a discontinuity at $t = N_{t}-1$ and the other choice will have a discontinuity
at the spatial boundary $x = N_{x}-1$. When using imaginary complex phase, it is possible to adjust the
value of the electric field such that the discontinuity disappears. However, these values are quite large for
the typical lattice sizes used in current studies. Moreover, this is only possible for imaginary electric fields
and it isn't clear that the results produced by such simulations can be analytically continued to real values
of the field.

In this study we will introduce the electric field using $A_{x} = Et$. This has a discontinuity at $t = N_{t}-1$ and
we will use Dirichlet boundary conditions in time to restrict quark propagation through the singularity. Even
with this choice, when using periodic boundary conditions in the spatial directions, in particular in the 
x-direction, there is an issue with regard to the origin of time~\cite{Engelhardt:2007ub}: in a infinite volume
the potentials $A_{x} = E(t-t_{0})$ lead to the same physical system, irrespective of the value of $t_{0}$.
Indeed, there is a gauge transformation that connects these potentials. This is no longer true for
a finite box with periodic boundary conditions. This can be easily seen if you consider the Polyakov loops
wrapping around the x-direction: the electric Polyakov loop for a given time-slice has different values for
different choice of $t_{0}$ and thus the two choices cannot be connected by a gauge transformation. Two
different choices of $t_{0}$ differ by a constant shift in the potential. When this constant potential is introduced
in the absence of an electric field, it can be expressed as a constant 
rotation for each link exactly as in the case of twisted boundary conditions~\cite{Bedaque:2004kc} and
we can think of the pion as a zero momentum combination of two quark/anti-quark fluxes running along 
the x-axis in opposite directions; the momentum of each flux is proportional to $1/L_{x}$. It is thus possible
that a change in $t_{0}$ could lead to a correction in the polarizability that vanishes with some power
of $L_{x}$ as the box size is increased. 

\begin{figure}[t]
\begin{center}
   \includegraphics[width=12cm]{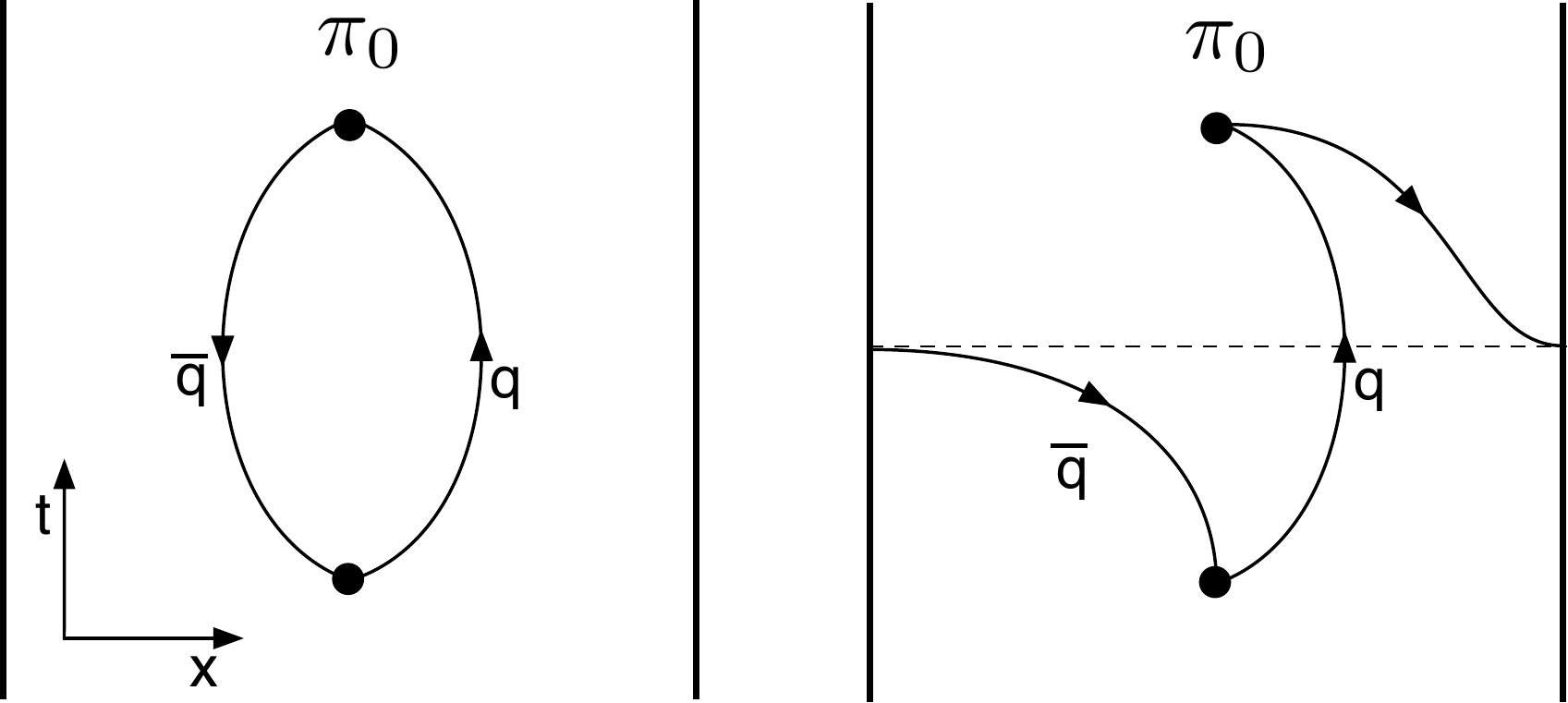} 
   \caption{Neutral pion contributions: the left panel indicates the $t_{0}$ invariant contributions and
   the right panel shows $t_{0}$ non-invariant contributions present when using periodic boundary
   conditions.
   \label{fig:2}}
\end{center}
\end{figure}

The $t_{0}$ dependence of the correlation function comes from the paths that wrap around the lattice in the 
x-direction 
(see Fig.~\ref{fig:2}). To remove this contributions we impose Dirichlet boundary conditions in the x-direction.
Together with Dirichlet boundary conditions in time, this makes the correlation functions for the neutral
particles completely gauge invariant even on a finite lattice. The change in the origin of time (or the origin
of space, $x_{0}$, if $A_{t}=-E(x-x_{0})$ potential is used) is now expressible
as a gauge transformation and the correlation functions are independent of the choice of origin. 
On the other hand, Dirichlet boundary conditions in space will force the quarks to acquire a
non-zero momentum of the order of $\pi/L_{x}$. This, in turn, can lead to finite volume effects of
the polarizability which will vanish only with some power of $1/L_{x}$ as we approach the continuum
limit.

Finally, we want to comment on the expectations regarding the ``connected'' contributions to the
neutral pion. The neutral pion interpolating field, $\pi_{0} = (\bar{u}\gamma_{5} u - \bar{d}\gamma_{5} d)/\sqrt{2}$, gives rise to a correlation function that has both connected and disconnected contributions. In
the isospin limit the disconnected contributions vanish. However, the presence of the electric field breaks
the isospin symmetry and the disconnected diagrams need to be included. Evaluating these diagrams is time
consuming and most polarizability studies disregard their contribution. The ``connected'' polarizability
is then evaluated using only the connected part of the correlation function. To better understand the 
behavior of the ``connected'' polarizability, it is important to realize that the correlation function corresponds
to an interpolating field of the form $\pi_{0}^{(c)} = \bar{u}\gamma_{5}u'$, where $u$ and $u'$ are two
distinct flavors of quarks that have the same charge. For such a theory, the ``pions'' are all uncharged and
a chiral perturbation theory ($\chi$PT) would predict that the electric field, to leading order, has no 
influence on hadrons. $\chi$PT predictions regard the effect of the electric field on the hadrons' pion cloud; 
the contribution measured by the ``connected'' polarizability is due to the deformation of the pions 
themselves. This contribution is neglected in $\chi$PT calculations.

\section{Numerical results}

As noted in the previous section, the finite volume corrections to polarizability are expected to be
dominated by powers of $1/L_{x}$. To gauge these corrections, we generated a set of
quenched ensembles with increasing size in the electric field direction: $N_{x}\times 24^{2}\times 48$,
with $N_{x}=24,36,48$. We used a Wilson gauge action with $\beta=6.0$ which corresponds to a 
lattice spacing of $a=0.093\fm$ and Wilson fermions since they have less problems with
exceptional configurations at small quark masses~\cite{Alexandru:2009id}. We computed the
polarizability for 4 values of the quark mass, $\kappa = 0.1567, 0.1565, 0.1562, 0.1546$,
 that correspond to a pion mass in the range 
$m_{\pi}=270-700\MeV$. For these quark masses $m_{\pi}L=3.2-8$, where $L=24 a$ is the smallest
spatial size, so that the finite volume effects, other than the ones discussed in the previous section, should
be small. The electric field, in dimensionless units $\eta=a^{2} |q_{d}| E = a^{2} e E/3 = 5.76\times 10^{-3}$, 
is set to the same value as in our previous study~\cite{Alexandru:2009id}. This electric field, while strong
when expressed in macroscopic units, was found to be within the range where the mass shift is still dominated
by the polarizability term. To extract the mass shift we use a correlated fit where both the zero field and 
non-zero field pion propagators are used simultaneously to allow us to take into account their 
cross-correlations~\cite{Alexandru:2009id}.

\begin{figure}[t]
\begin{center}
   \includegraphics[width=12cm]{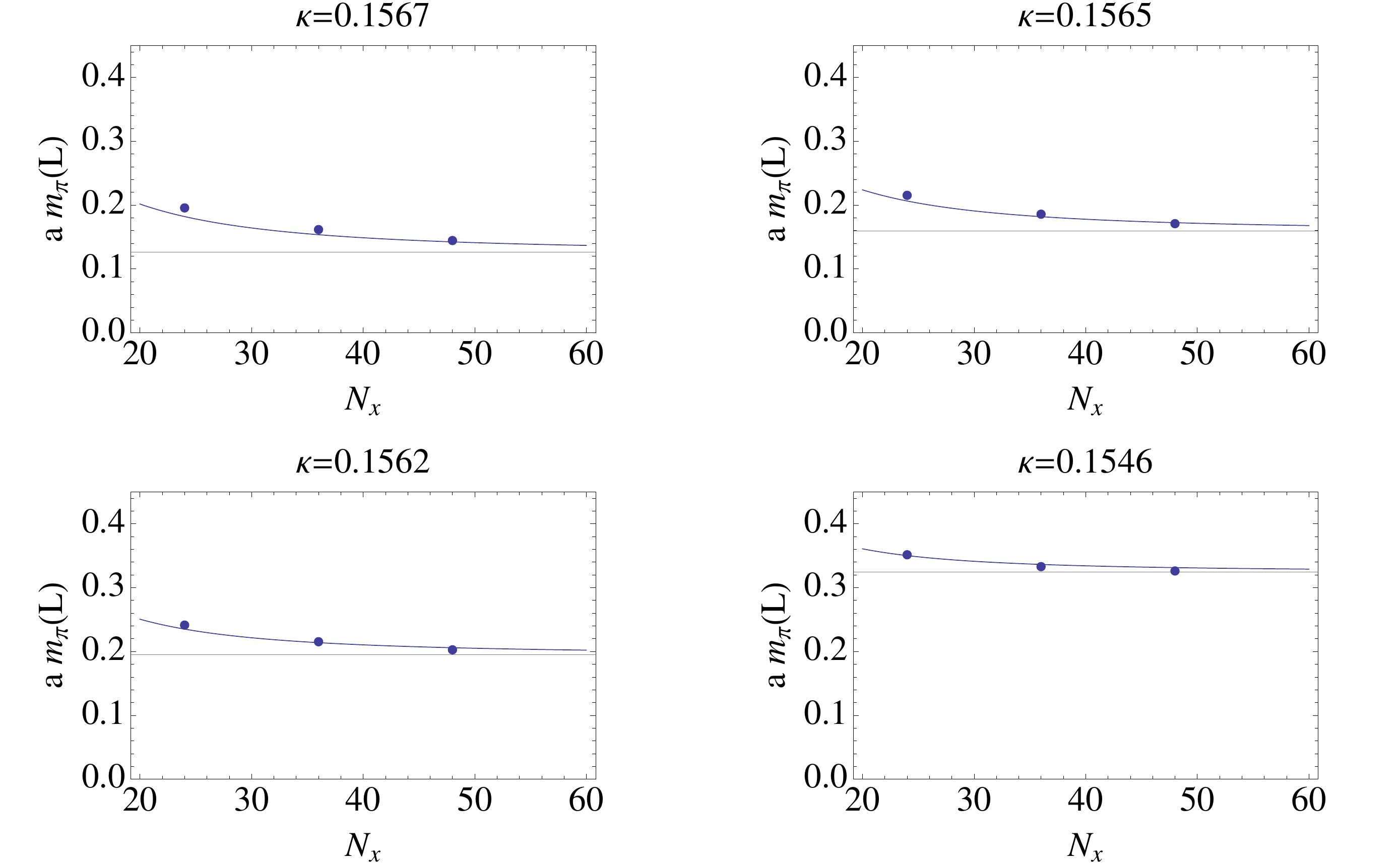} 
   \caption{Lowest energy state in the pion channel on lattices with Dirichlet boundary conditions. The
   continuous curve is the expected energy for a relativistic particle with mass $m_{\pi}$ and momentum
   $\pi/L$.
      \label{fig:3}}
\end{center}
\end{figure}

It is worth pointing out here that the energy of the lowest pion state 
is modified by the presence of the Dirichlet boundary
conditions in the x-direction. The reason for this is the fact that the quark fields are forced to vanish at the
boundary which adds a non-zero momentum to the quark fields; the energy of the pion is thus shifted. 
Treating the pion as a point particle confined to a box of size $L_{x}$, basic quantum mechanics would
predict that $m_{\pi}(L)=\sqrt{m_{\pi}^{2} + (\pi/L)^{2}}$. In Fig.~\ref{fig:3} we plot the lowest energy in the
pion channel and compare it with the above prediction. The pion mass, $m_{\pi}$, is computed on the 
$24^{3}\times 48$ lattices using periodic boundary conditions.

\begin{figure}[t]
\begin{center}
   \includegraphics[width=7.5cm]{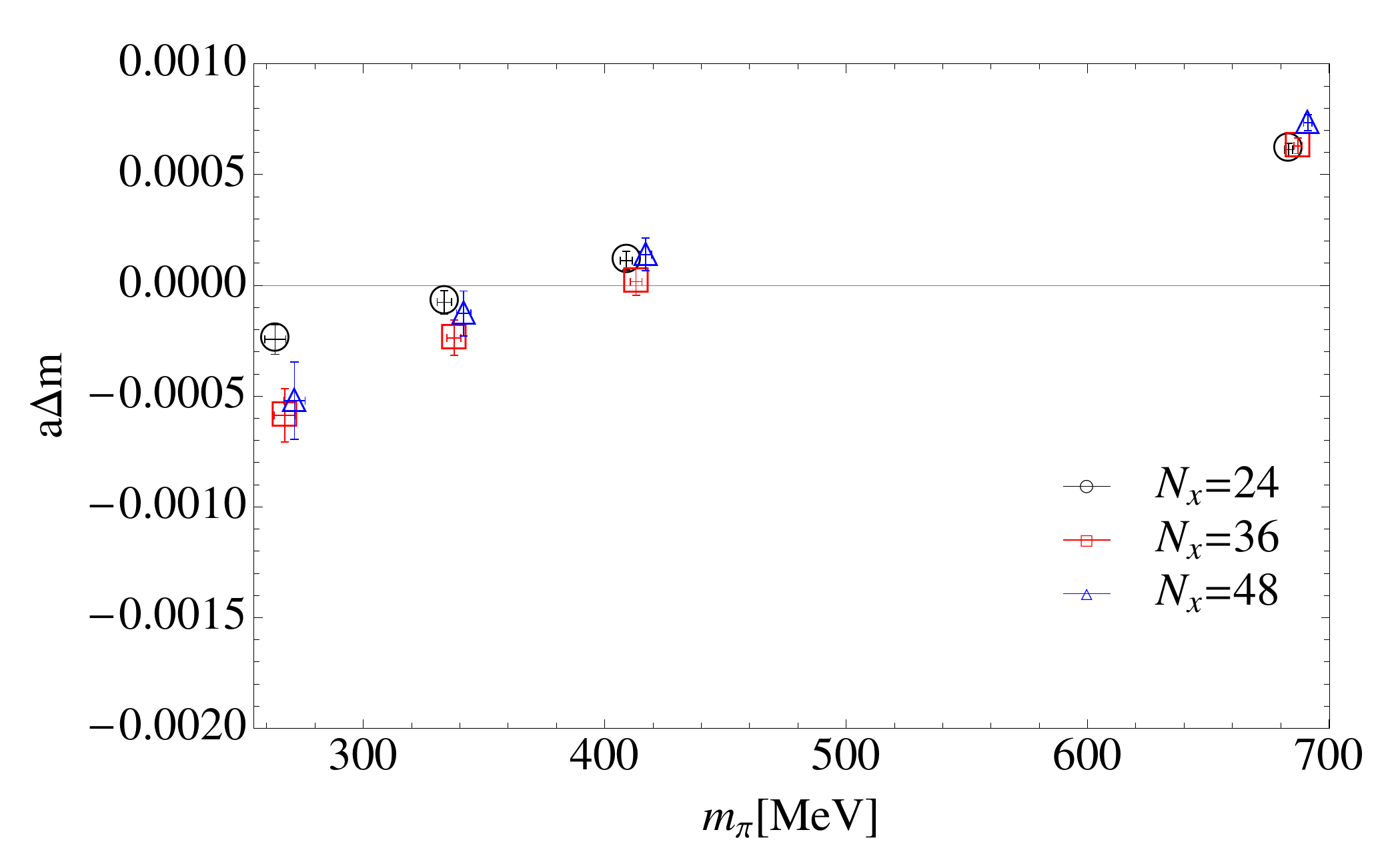} 
    \includegraphics[width=7.5cm]{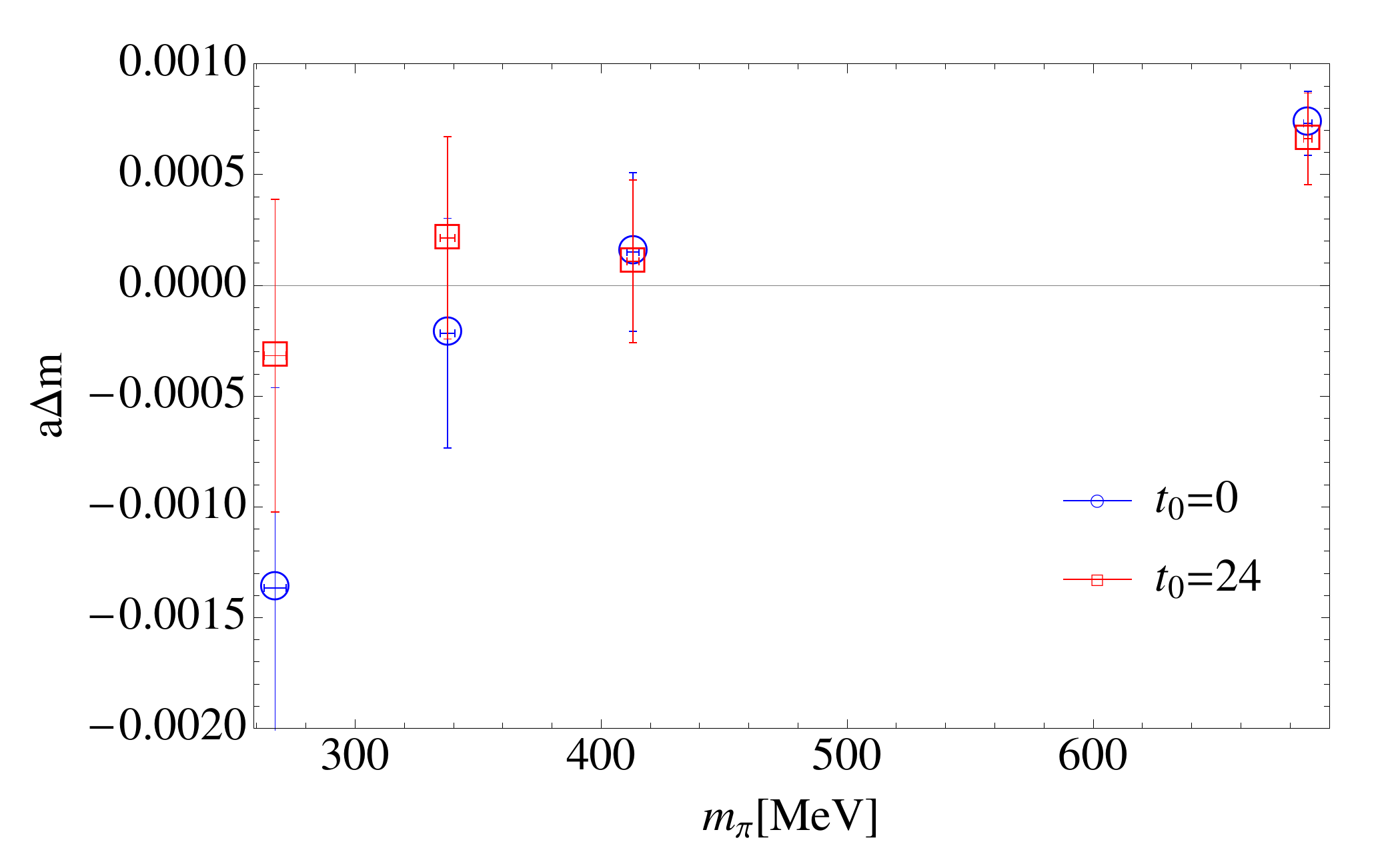} 
   \caption{Left: Mass shift induced by the electric field for the Dirichlet boundary conditions case. 
   The data points are displaced in the horizontal direction by $\pm 4\MeV$ for better visibility.
   Right: The same mass shifts measured on $24^{3}\times 48$ lattice using periodic boundary
   conditions.
      \label{fig:4}}
\end{center}
\end{figure}

In the left panel of Fig.~\ref{fig:4} we present the mass shift for the pion calculated using
Dirichlet boundary conditions for each of the three lattice sizes and in the right panel
the results for the mass shift computed using periodic boundary conditions. In the periodic
case, we have used $A_{x} = E(t-t_{0})$ with $t_{0}=0$ and $t_{0}=24$. While the results differ 
for the two choices of the origin of time, they agree within error bars. Moreover, given the
rather large error bars for the periodic case, the Dirichlet boundary conditions results are
also compatible with them. The size of the error bars in the periodic case is a bit surprising. 
It can only partly be explained by the difference in statistics: 1000 configurations for the Dirichlet
boundary compared to 600 for the periodic case. To draw any definitive conclusions, we need
to increase the statistics for the periodic case in order to reduce the size of the error bars.

\begin{figure}[t]
\begin{center}
   \includegraphics[width=12cm]{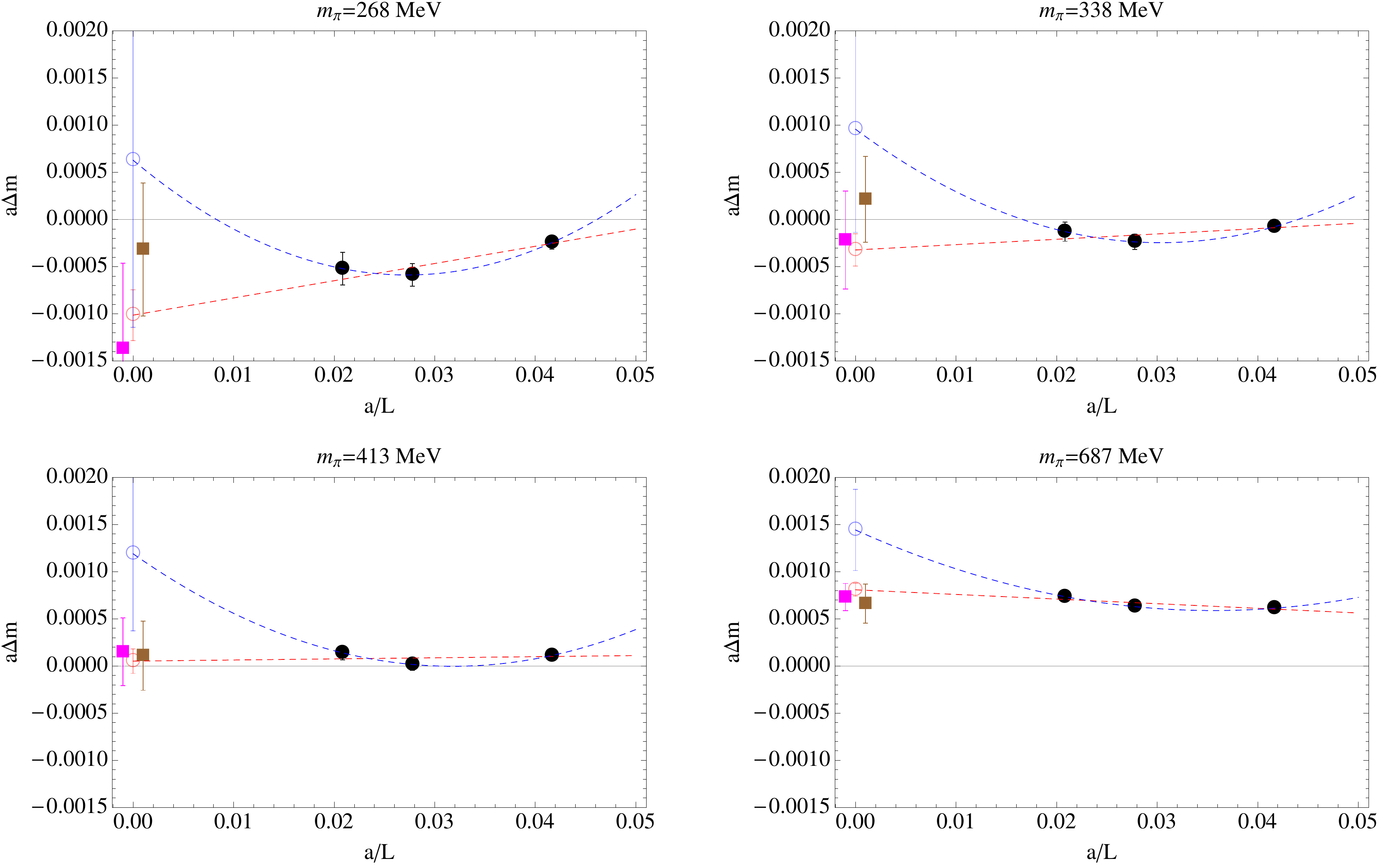} 
   \caption{Infinite box extrapolation: the dotted lines represent the linear
   and quadratic fits, the black points are the Dirichlet results and the red and blue squares
   are the results for $24^{3}\times48$ boxes with periodic boundary conditions in the spatial directions.
      \label{fig:5}}
\end{center}
\end{figure}

For pion masses lower than $400\MeV$
the mass shift becomes negative, as seen in our previous study. The mass shifts vary as we change 
the lattice size. Focusing on the region where the mass shift is negative, we see that the change
is not monotonic. To get a better sense of the trend we need to focus on each mass individually and
perform an extrapolation. Unfortunately, none of the analytical calculations predict a functional form
for the finite volume corrections of this quantity. As discussed in the previous section, we expect the
correction to be power-like in $1/L_{x}$. In Fig.~\ref{fig:5} we show the results of two
extrapolation based on linear and quadratic forms for the finite volume corrections. 
Both forms fit the data reasonably well and we cannot use the quality of the fit to determine
which form is more suitable. The linear extrapolation predicts negative polarizabilities even in the 
infinite volume limit ($L_{x}\rightarrow\infty$), whereas the quadratic form extrapolates to a positive polarizability.
While the linear extrapolation has smaller errors and it agrees well with the results obtained
from periodic boundary conditions, the large error bars for the quadratic fit also show reasonable
agreement with periodic data. 

\section{Conclusions and outlook}

We have computed the mass shift for the neutral ``pion'' with both Dirichlet and periodic boundary conditions.
For Dirichlet case, we have used ensembles with three different lattice sizes and we found that at
lower pion masses the polarizability stays negative even on larger lattices. We cannot reliably
extrapolate to infinite volume partly because we are lacking an analytical prediction and partly
because of the limited quality of our data. We plan to add one more ensemble to this study in order
to increase the reliability of our extrapolation.

The mass shifts computed using periodic boundary conditions agree within error bars with the ones
computed using Dirichlet boundary conditions. Moreover, we looked at the dependence of these 
results on the choice of the origin of time and we found that while the mass shift changes, the change
is comparable with the error bars. We have to stress that the error bars in the periodic case were
significantly larger and some of the conclusions might change when we reduce the errors.
We plan to increase the number of configurations in our ``periodic'' ensemble and carry out
simulations on larger lattices to gauge the finite volume effects.

\section{Acknowledgements}

The authors are supported in part by U.S. Department of Energy under grant DE-FG02-95ER-40907. 
The computational resources for this project were provided in part by the George Washington University 
IMPACT initiative.

\end{document}